\documentclass[a4paper]{article}

\usepackage{graphicx}
\usepackage{url}

\title{Reflectivity and Exact Asymptotic Forms\\ of the Dielectric Function of Non-Ideal Plasmas}

\author{D. Ballester, A.M. Fuentes, I.M. Tkachenko\\
{\it Universitat Polit\`ecnica de Val\`encia, 46022 Valencia, Spain}}

\begin{document}

\maketitle

\begin{abstract}
The problem of calculation of the reflectivity of non-ideal shock-com\-pressed plasmas is revisited. The dielectric formalism based on the method of moments incorporating exact asymptotic forms and sum rules is applied to the new experimental data. The approach is based on the \textit{Ansatz} of reflection of laser radiation from a thin shock-front and possesses one adjustable parameter which depends only on the thermodynamic conditions of the experiment. A self-consistent agreement with all available experimental data \cite{exp} is achieved.
\end{abstract}

\section{Introduction}
In general, in the long-wavelength limit, the real part of the internal dynamic conductivity of dense multicomponent plasmas only possesses two even convergent moments \cite{ATs}, whereas odd ones vanish due to its parity (obviously, in absence of any external magnetic field). By means of the Nevanlinna formula, the (complex) internal conductivity can be recast as (see \cite{ATs} and references therein for details):





\begin{equation}
\sigma\left(\omega\right)=\frac{i\omega_{p}^{2}}{4\pi}\frac{\omega+q\left(\omega\right)}{\omega^{2}+\Omega^{2}+\omega q\left (\omega\right)},\label{Nev}
\end{equation} where $\omega_{p}$ is the plasma frequency. In addition, $\Omega^{2}$ is given by the quotient between the second moment of the conductivity and the $f$-sum rule, whereas the model parameter function $q(z)$ is a Nevanlinna-class function such that $\lim_{z\rightarrow\infty}\left(q(z)/z\right)=0$ within any sector of the upper-half plane $\delta<\arg(z)<\pi-\delta$, $0<\delta<\pi/2$.

\section{Modelling of $q(z)$ and the asymptote}

In order to reproduce the well-known Perel' and Eliashberg \cite{PE} asymptote in a completely ionized plasma, we propose the following model parameter function for (\ref{Nev}):
\begin{equation}
q\left(\omega\right)=C\left({\rm sign}\left(\omega\right)\sqrt{\left\vert\omega\right\vert}+i\sqrt{\left\vert\omega\right\vert}\right)-\frac{\zeta\omega_{p}\Omega^{2}\tau}{z+i\zeta\omega_{p}}, \label{q}
\end{equation} where $C=2^{1/2}3^{-5/4} r_{s}^{3/2} \omega_{p}^{5/2} \Omega^{-2}$, $r_{s}$ the Brueckner parameter, and $\zeta>0$ is a model parameter.

Next, the model parameter $\zeta $ in (\ref{q}) has been used to fit the experimental data of \cite{exp}. To this aim, the well-known long-wavelength Fresnel formula has been applied. In addition, the values of the static conductivity have been taken either from \cite{exp} or extrapolated for similar thermodynamical conditions, whereas in order to calculate the corresponding electron and ion densities we have assumed in all cases appearing in \cite{exp} that the ionization degree was equal to unity. Further, calculations show that the model parameter $\zeta$, taking into account the precision of determination of plasma static characteristics, varies consistently in the realm of thermodynamic conditions studied in \cite{exp}.

\begin{table}[h]
\begin{center}
\caption{\label{table1} Fit of the parameter $\zeta$ in expression (\ref{q}) corresponding to the reflectivity experimental data obtained by \cite{exp} for a $Xe$ plasma (light source with $\lambda =532$ $nm$).}
\begin{tabular}{cccc|cc}
\hline \multicolumn{4}{c|}{\small Experimental data} & \multicolumn{2}{c}{\small Model}
\\ \hline
\small $p(GPa)$ & \small $T(K)$ & \small $\rho(g\cdot cm^{-3})$ & \small $R$ & \small $\zeta$ & \small $R$
\\ \hline
\small 4.1 & \small 33100 & \small 1.1 & \small 0.02 & \small 0.40 & \small 0.06
\\
\small 6.1 & \small 33120 & \small 1.6 & \small 0.045 & \small 0.29 & \small 0.065
\\
\small 9.1 & \small 32090 & \small 2.2 & \small 0.10 & \small 0.42 & \small 0.10
\\
\small 12.0 & \small 32020 & \small 2.8 & \small 0.16 & \small 0.55 & \small 0.16
\end{tabular}
\end{center}
\end{table} \normalsize

\begin{table}[h]
\begin{center}
\caption{\label{table2} Fit of the parameter $\zeta$ in expression (\ref{q}) corresponding to the reflectivity experimental data obtained by \cite{exp} for a $Xe$ plasma (light source with $\lambda =694$ $nm$).}
\begin{tabular}{cccc|cc}
\hline \multicolumn{4}{c|}{\small Experimental data} & \multicolumn{2}{c}{\small Model}
\\ \hline
\small $p(GPa)$ & \small $T(K)$ & \small $\rho(g\cdot cm^{-3})$ & \small $R$ & \small $\zeta$ & \small $R$
\\ \hline
\small 0.93 & \small 32070 & \small 0.27 & \small 0.02 & \small 0.437 & \small 0.02
\\
\small 1.9 & \small 32900 & \small 0.53 & \small 0.05 & \small 0.477 & \small 0.05
\\
\small 4.1 & \small 33100 & \small 1.1 & \small 0.11 & \small 0.424 & \small 0.11
\\
\small 6.1 & \small 33120 & \small 1.6 & \small 0.14 & \small 0.367 & \small 0.14
\\
\small 9.1 & \small 32090 & \small 2.2 & \small 0.18 & \small 0.385 & \small 0.18
\\
\small 12.0 & \small 32020 & \small 2.8 & \small 0.26 & \small 0.512 & \small 0.26
\\
\small 16.0 & \small 31040 & \small 3.4 & \small 0.40 & \small 1.73 & \small 0.40
\end{tabular}
\end{center}
\end{table} \normalsize

\begin{table}[h]
\begin{center}
\caption{\label{table3} Fit of the parameter $\zeta$ in expression (\ref{q}) corresponding to the reflectivity experimental data obtained by \cite{exp} for a $Xe$ plasma (light source with $\lambda =1064$ $nm$).}
\begin{tabular}{cccc|cc}
\hline \multicolumn{4}{c|}{\small Experimental data} & \multicolumn{2}{c}{\small Model}
\\ \hline
\small $p(GPa)$ & \small $T(K)$ & \small $\rho(g\cdot cm^{-3})$ & \small $R$ & \small $\zeta$ & \small $R$
\\ \hline
\small 1.6 & \small 30050 & \small 0.51 & \small 0.096 & \small 0.30 & \small 0.096
\\
\small 3.1 & \small 29570 & \small 0.97 & \small 0.12 & \small 0.22 & \small 0.12
\\
\small 5.1 & \small 30260 & \small 1.46 & \small 0.18 & \small 0.20 & \small 0.18
\\
\small 7.3 & \small 29810 & \small 1.98 & \small 0.26 & \small 0.23 & \small 0.26
\\
\small 10.5 & \small 29250 & \small 2.70 & \small 0.36 & \small 0.33 & \small 0.36
\\
\small 16.7 & \small 28810 & \small 3.84 & \small 0.47 & \small 2 & \small 0.45
\end{tabular}
\end{center}
\end{table} \normalsize

\end{document}